\documentstyle[prl,aps]{revtex}
\input BoxedEPS.tex
\SetepsfEPSFSpecial
\HideDisplacementBoxes

\begin{document}

\twocolumn[\hsize\textwidth\columnwidth\hsize
          \csname @twocolumnfalse\endcsname
\title{Possible superconductivity above 400 K in carbon-based multiwall 
nanotubes}
\author{Guo-meng Zhao$^{*}$ and Y. S. Wang}
\address{
Department of Physics and Texas Center for Superconductivity, 
University of Houston, Houston, Texas 77204, USA}

\maketitle
\widetext
\vspace{0.3cm}
\maketitle
\widetext

\begin{abstract}

Magnetization and resistance measurements were carried out on 
carbon-based  multiwall nanotubes. Both magnetization and resistance data 
can be consistently explained in terms of bulk superconductivity above 
400 K although we cannot completely rule out other possible explanations 
to the data.
~\\
~\\
\end{abstract}
]
\narrowtext

Seven years  after the discovery of the 30 K superconductivity  in 
a single-layer copper-based oxide \cite{KAM86}, the superconducting transition temperature 
$T_{c}$ was  raised up to 153 K in a mercury-based three-layer cuprate under a 
pressure of 17 kbar \cite{Chu}. Meanwhile, high-temperature superconductivity 
as high as 40 K was observed in electron-doped C$_{60}$ (see a review 
article \cite{Gunnarsson}). Last year, the superconductivity at 52 K was observed in 
hole-doped C$_{60}$ \cite{Batlogg1}. Very recently, a maximum $T_{c}$ = 117 K has 
been discovered in a hole-doped C$_{60}$/CHBr$_{3}$ \cite{Batlogg2}. It was also 
suggested that \cite{Batlogg2} the maximum $T_{c}$ might reach up to 150 K if one 
could expand spacing between the C$_{60}$ molecules further. Here we 
report magnetization and resistance measurements on 
carbon-based  multiwall nanotubes. Both magnetization and resistance data 
can be consistently explained in terms of bulk superconductivity above 
400 K although we cannot completely rule out other possible explanations 
to the data.

The commercially available multiwall nanotubes were prepared from the high-purity 
graphite (99.9995$\%$) by an arc process with no metal 
catalysts. The nanotubes consist of 5-20 graphite layers with 2-20 nm in diameter and 
100 nm to 2~$\mu$m in length. The tubes  are normally assembled 
into a bundle, and the bundles are assembled into ropes.  We selected 
several ropes
with a diameter of 0.05-0.1 mm and a length of  0.5-1 mm 
for resistivity measurements. The four point contacts were attached by 
conductive silver epoxy and the contact resistance is in the range of 
20-60 $\Omega$. Magnetization was measured by a quantum design superconducting
quantum interference device (SQUID). We can attain high accuracy in the measurements
by using the reciprocating sample option (the accuracy is better 
than 3 $\times$ 10$^{-8}$ emu). The SQUID response is well centered at 
each temperature. The ``zero-field'' of less than 3 mG was achieved by 
using the ultra-low field option. The magnitude and direction of 
an applied magnetic field were determined by comparing the 
diamagnetic signal of a standard Pb superconductor.

In Fig.~1a, we show temperature dependence of the field-cooled (FC) 
susceptibility in the magnetic fields of 20 mG and 280 mG for sample 
ZWC, which has a weight of 165 mg. The sample 
was cooled in the fields from 400 K. The sample did not experience 
any high magnetic fields before these measurements. It is apparent 
that the diamagnetic susceptibility persists up to 400 
K; the diamagnetic signal at 5 K corresponds to  about 0.2$\%$ of the 
full Meissner effect. It is remarkable that the diamagnetic susceptibility strongly depends 
on the magnetic field.  Moreover, there is a kink 
feature at $\simeq$ 110 K, which may be related to the 
intergrain Josephson coupling as discussed below.
\begin{figure}[htb]
\input{epsf}
\epsfxsize 7cm
\centerline{\epsfbox{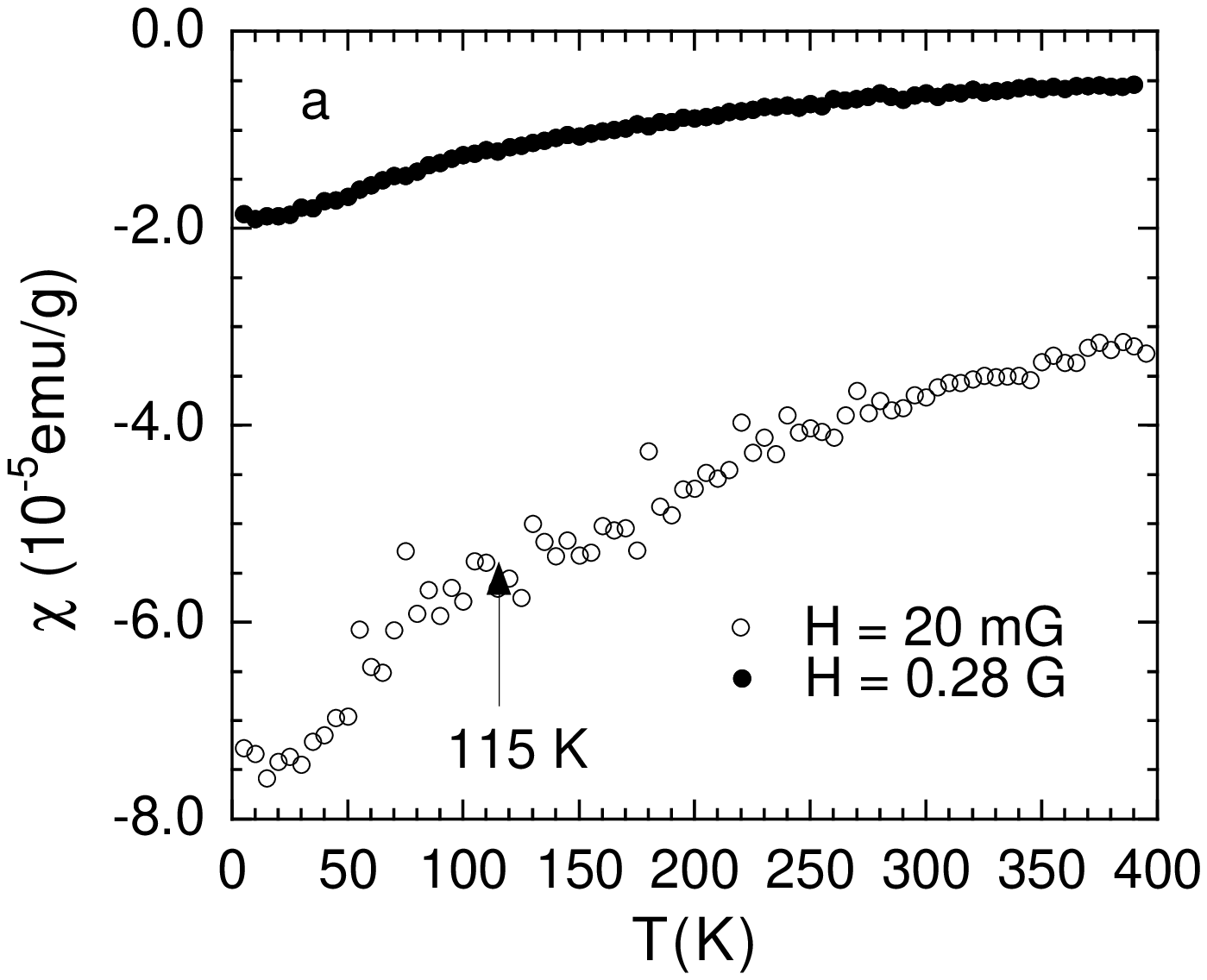}}
\vspace{0.4cm}
\input{epsf}
\epsfxsize 7cm
\centerline{\epsfbox{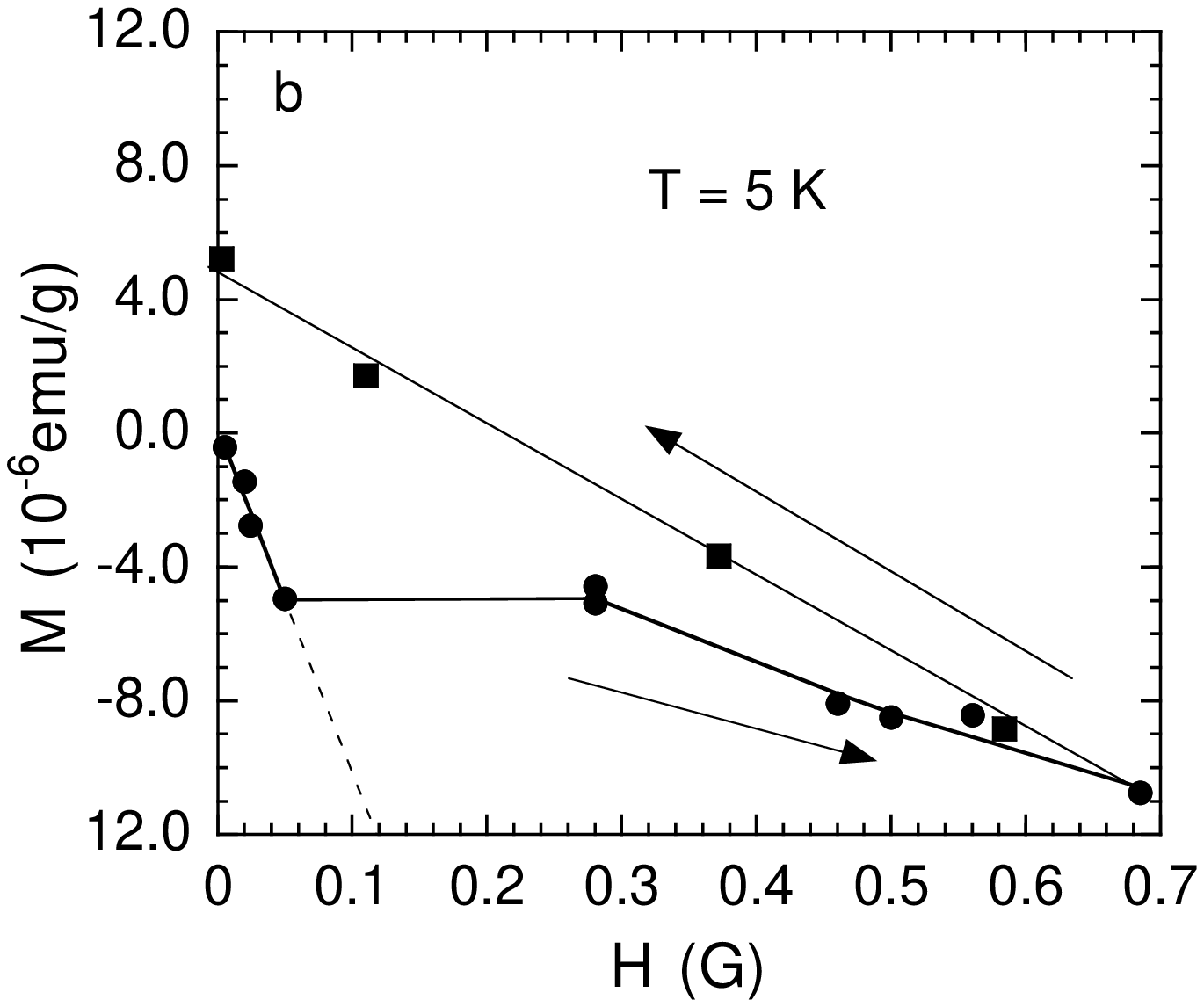}}
	\vspace{0.6cm}
	\caption {a) Temperature dependence of the field-cooled (FC)
	susceptibility for sample ZWC (168 mg) in the magnetic fields of 20 mG and 280 mG.  b) The FC magnetization as a function of 
	field in the low-field range.}
	\protect\label{fig1}
\end{figure}
The strong field dependence of the diamagnetic susceptibility 
cannot arise from the core 
diamagnetism of the carbon ions and/or from the orbital diamagnetism 
of conduction electrons.   In addition, the moment of 
our sample holder is linearly 
proportional to the field and has a magnitude of $-1.9\times 10^{-5}$ emu 
in the field of 0.1 T.  The observed strong field 
dependence of the diamagnetic susceptibility in this material is  
similar to that in single crystalline and polycrystalline samples of 
the cuprate superconductors  \cite{Malozemoff,Tomioka}.

In Fig.~1b, we display magnetization as a function of the field in the 
low field range. In the very low field range, the magnetization is linearly 
proportional to the field, as expected for a superconductor. Furthermore, the 
detailed field 
dependence of the magnetization at 4.2 K from the earth field  to 10 T was 
reported in Ref.~\cite{Tsebro}. The field dependence of the 
magnetization in the low field 
range (Fig.~1b) and in the high field range ~\cite{Tsebro} 
is similar to that for a heavily overdoped 
Tl$_{2}$Ba$_{2}$CuO$_{6+y}$ in a temperature slightly below $T_{c}$ ($\sim$ 
15 K) \cite{Berg}. A possible theoretical explanation to the unusual field 
dependence of the magnetization can be found in Ref.~\cite{Millis}.

Fig.~2 shows the field-cooled and zero-field cooled 
(ZFC) susceptibility. For the ZFC measurement, the sample was cooled 
from 400 K to 5 K in a field of less than 3 mG, and then a field of 280 
mG was set. After the measurement was finished, the field was precisely 
determined by comparing the 
diamagnetic signal of a standard Pb superconductor. It is clear that the difference between the ZFC 
and FC signals  is significant (about 15 $\%$ 
of the FC signal at 5 K). This behavior is also expected 
from a superconductor, as seen in cuprates where the difference 
between the ZFC and FC signals  is  about 20-30$\%$ of the FC 
signal \cite{Malozemoff}. 
\begin{figure}[htb]
\input{epsf}
\epsfxsize 7cm
\centerline{\epsfbox{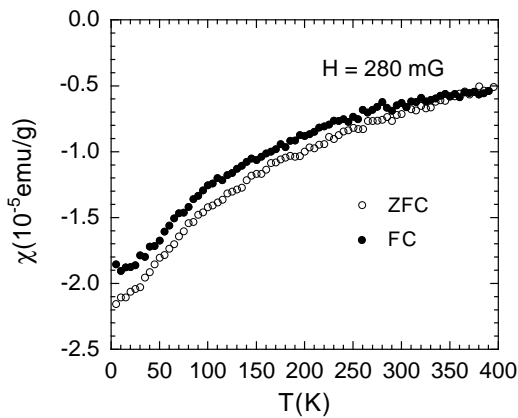}}
\vspace{0.4cm}
	\vspace{0.6cm}
	\caption {Temperature dependence of field-cooled and zero-field cooled 
(ZFC) susceptibility. It is clear that the difference between the ZFC 
and FC signals  is significant (about 15$\%$ 
of the FC signal at 5 K).}
	\protect\label{fig2}
\end{figure}
Fig.~3  plots the remnant magnetization as a function of temperature. 
A magnetic field of 5 T was applied at 400 K. After the sample was cooled 
from 400 K to 5 K under the field, we set the field to zero (the real 
field is -1.1G) and measured the magnetization from 5 K to 400 K. One can 
clearly see that the temperature dependence 
of the $M_{r}$ is similar to that of the diamagnetic susceptibility 
in a field of 20 mG except for the opposite signs  (see Fig.~1a and 
Fig.~3). This behavior is 
expected for a superconductor, as seen in 
cuprates \cite{Malozemoff}. If there were a mixture of a 
ferromagnet and a material with field independent diamagnetic 
susceptibility, the total susceptibility would tend to turn up 
below 120 K where the $M_{r}$ increases suddenly. In contrast, the 
low-field susceptibility suddenly turns down rather than turns up below 
120 K. This provides strong evidence that the observed $M_{r}$ in the 
nanotubes is intrinsic and has nothing to do with the presence of 
ferromagnetic impurities.  In addition, by comparing the magnitude of 
the $M_{r}$ for the present sample with that for the sample in 
Ref.~\cite{Tsebro}, one can see that the remnant 
moment $M_{r}$ is nearly proportional to the magnitude of the diamagnetic 
susceptibility. This gives further support to the explanation that the 
$M_{r}$ is related to the diamagnetism rather than to the presence of 
ferromagnetic impurites. The fact that the $M_{r}$ 
remains up to 400 K implies that the bulk superconductivity might persist up to 400 
K. 

\begin{figure}[htb]
\input{epsf}
\epsfxsize 7cm
\centerline{\epsfbox{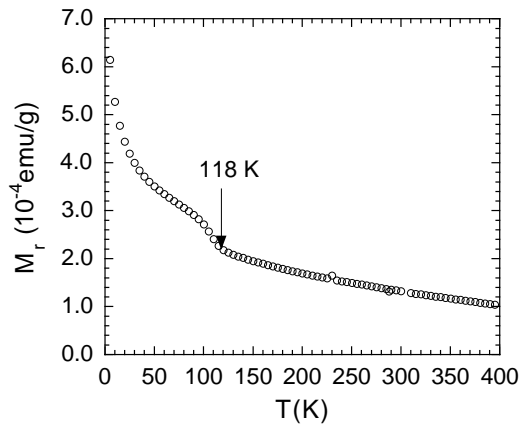}}
\vspace{0.4cm}
	\vspace{0.6cm}
	\caption {The remnant magnetization as a function of 
	temperature for sample ZWC. 
	A magnetic field of 5 T was applied at 400 K. After the sample 
	was cooled from 400 K to 5 K under the field, the field was set to zero 
	and  the magnetization was measured from 5 K to 400 K.}
	\protect\label{fig3}
\end{figure}

It is known that the magnitude of $M_{r}$ in a superconductor  is proportional to 
the critical 
current density $J_{c}$. The fact that the magnitude of $M_{r}$ is 
rather small and decreases rapidly with increasing temperature 
suggests that the critical current is limited by intergrain Josephson coupling. 
There is a pronounced kink feature at 
$\simeq$ 120 K in the 
diamagnetic susceptibility curve (see Fig.~1a), and in the  $M_{r}$ 
curve (see Fig.~3a). This feature may be related to the Josephson coupling 
among the bundles.  There may be a stronger 
intertube Josephson coupling within the bundles, which 
may persist up to 400 K in the very low fields. This can naturally 
explain why the low-field (20 mG) susceptibility is substantially higher than 
the ``high-field''  ($>$ 0.28 G) susceptibility for $T <$ 400 K. It 
appears that a field of $>$ 0.28 G is strong enough to suppress the 
Josephson coupling.  

In Fig.~4a, we present the resistance data from 300 K to 750 K for 
sample ZW-8. The sample was measured in flowing Ar 
gas to avoid oxidation. The solid black line is the fitted smooth curve below 550 
K.  It is remarkable that the resistance suddenly turns up above 600 K, 
which may manifest the resistive transition to the normal state within 
the individual superconducting tubes although we cannot completely 
rule out other possibilities. Such a temperature dependence of the 
resistance is reproducible in all the three samples in warming up 
measurements. 
However, the resistance becomes lower and the resistance jump nearly 
disappears for the cooling down measurements. This could be understood as 
follows. If $T_{c}$ is widely distributed, the resistive path could be 
shortened by a superconducting path with higher $T_{c}$ for 
cooling down measurements  (i.e., the current path is possibly 
altered for the cooling down and warming up measurements). 

The non-zero resistance 
below $T_{c}$ might be due to the fact 
that the superconducting 
fraction $f_{s}$ in the ropes is not large enough to reach a percolation 
limit. Another possible reason is that the 
transport has 1-dimensional nature, so that the possibility for the short 
tubes (the average length is about 1~$\mu$m) to be connected into 
``long'' tubes of about 1 mm (through Josephson coupling) is extremely 
low. This has been indeed demonstrated in the transport measurements 
on the ropes of single-wall nanotubes, which show that there exists 
only one ``long'' tube in a 
rope \cite{Bockrath}. In worse cases, the intrinsic metallic 
conductivity in nanotube ropes can be completely masked by the 
intertube contact resistance and the resistance contributed from 
semiconductive tubes. The semiconducting behavior 
of the rope will still remain when some individual short tubes become 
superconducting. Even if there possibly exist 
several ``long'' tubes, the current density within 
the small number of the ``long'' tubes may exceed the Josephson 
critical current. Only in one of 14 ropes we measured, we found that 
the resistance at 5 K decreases with current when the measuring 
current is less than 10 nA. 
However, the result could not be reproduced for the second measurement 
on the same rope possibly because the very weak link was already broken.

In order to see more clearly the resistive transition, we plot in 
Fig.~4b the difference $R(T)$ - $R_{fit}(T)$, where 
\begin{figure}[htb]
    \input{epsf}
\epsfxsize 7cm
\centerline{\epsfbox{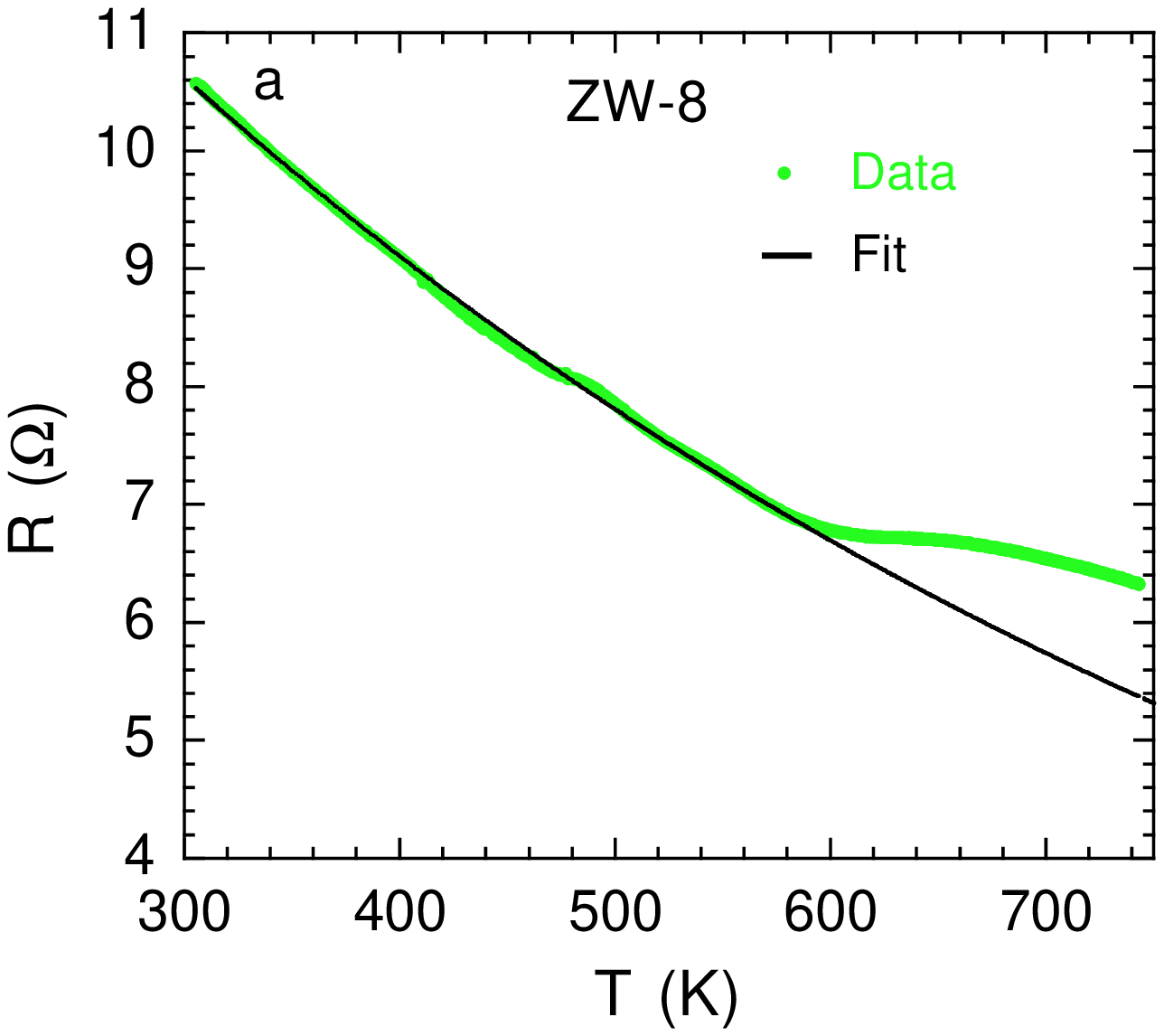}}
\vspace{0.4cm}
	\input{epsf}
\epsfxsize 7cm
\centerline{\epsfbox{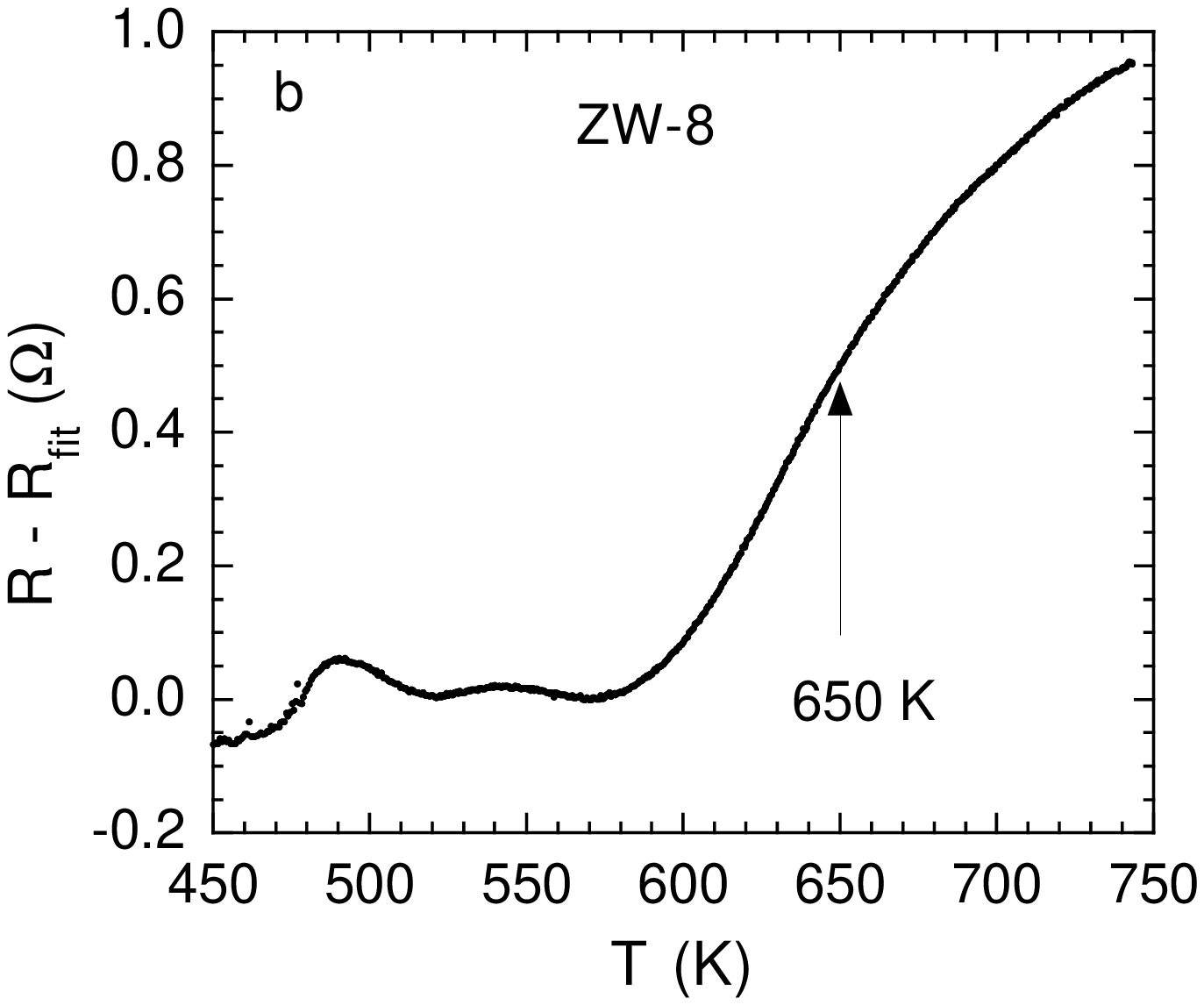}}
\vspace{0.4cm}
\input{epsf}
\epsfxsize 7cm
\centerline{\epsfbox{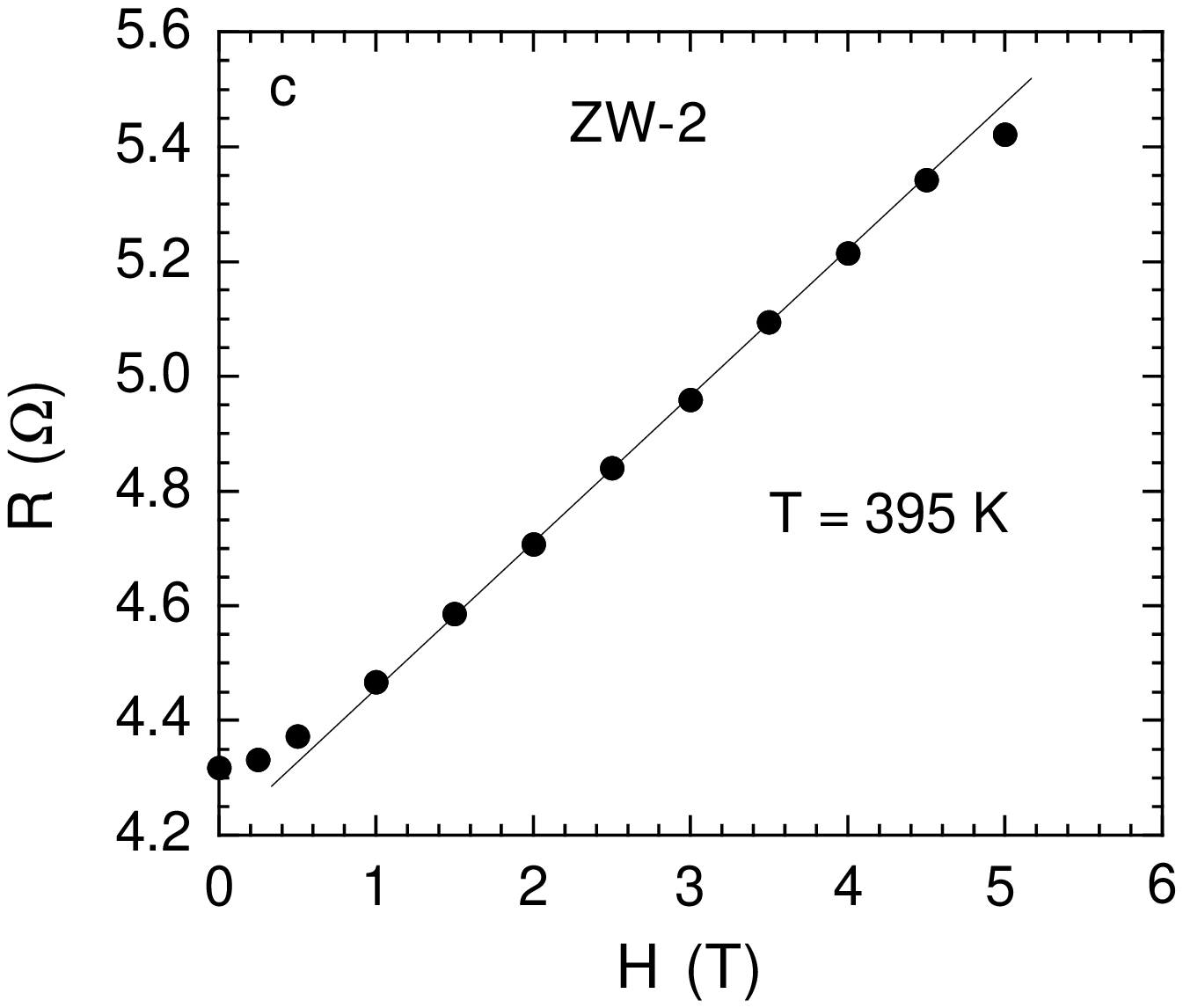}}
	\vspace{0.6cm}
	\caption {a) The resistance data from 300 K to 750 K for 
sample ZW-8. The sample was measured in flowing Ar 
gas to avoid oxidation.  The solid black line is the fitted smooth curve below 550 
K. b) The difference between $R(T)$ and $R_{fit}(T)$ ($R_{fit}(T)$ is 
the fitted black line in Fig.~3a). c) The magnetoresistance (MR) effect 
for sample ZW-2 at 395 K. }
	\protect\label{fig4}
\end{figure}
\noindent
$R_{fit}(T)$ is 
the fitted curve in Fig.~4a. The transition appears 
rather broad and may be incomplete even at 750 K. The resistance jump 
at $T_{c}$ is on the order of 1 $\Omega$. This is in good agreement 
with the magnetoresistance effect at 395 K, as shown in Fig.~4c.  The 
resistance was measured when the field was applied perpendicular to the 
axis of the rope. It 
is clear that there is a large positive magnetoresistance (MR) effect 
for the field between 0.5 T and 4.5 T. For the field below 0.5 T and above 
4.5 T, the 
MR effect appears saturated. This behavior is also expected from a 
superconductor (for an example, see Fig.~5 of Ref.~\cite{Affronte}). 
It is striking that the 
resistance jump from the superconducting to the normal state is also on 
the order of 1 $\Omega$, in agreement with the result shown in Fig.~4b. 
This consistency implies that the resistance jump at about 650 K 
should be related to the superconducting transition. Then the normal 
state resistivity should be 
on the order of 1000 $\mu\Omega$cm. Considering the correction for the porosity 
of the ropes \cite{Lee}, the normal state resistivity of the 
superconducting tubes would be on the order 
of 20 $\mu\Omega$cm, which is comparable with that ($\sim$34 $\mu\Omega$cm) 
for the metallic single-wall nanotubes \cite{Thess}.

Now we turn to discussion of the Meissner effect. For decoupled superconducting grains,  the 
diamagnetic susceptibility is given by \cite{Millis}
\begin{equation}
\chi (T) = -f_{s}\frac{\bar{R}^{2}}{40\pi\lambda^{2}(T)},
\end{equation}
where $\bar{R}$  is the average radius of spherical grains and 
$\lambda (T)$ is the grain penetration depth, which must be larger 
than $\bar{R}$ in order for Eq.~1 to be valid. In the present case, the 
grains are not spherical but cylindric. We may replace 
$\bar{R}$ by $\bar{r}$, and 1/40 by 1/60 (due to a difference in the 
demagnetization factor) when  the measuring field is along the tube axis 
direction, where $\bar{r}$
is the average radius of the tubes.  With 
$\bar{r}$= 50~\AA~\cite{Chau}, $f_{s}$ = 0.15, weight density of  
2.17 g/cm$^{3}$ \cite{Qian}, and $\chi_{\parallel} (0) = -1.1\times 10^{-5}$ 
emu/g \cite{Chau} (we have ignored the diamagnetic contribution in 
the normal state, which should be about $-5\times 10^{-7}$ emu/g 
according to a simple rolling model of graphite sheets), we find 
that $\lambda_{\parallel} (0)/\bar{r}$ = 5.8, and $\lambda_{\parallel} (0)$ 
$\simeq$ 289 \AA. If we assume $f_{s}$ = 1, then $\lambda_{\parallel} (0)$ 
$\simeq$ 746 \AA, which is also quite reasonable. Now if we take $\lambda_{\parallel} (0)$ 
$\simeq$ 1600 \AA (close to that for optimally doped cuprates), and 
$f_{s}$ = 1, we calculate $\chi_{\parallel} (0) = -2.4 \times 10^{-6}$ 
emu/g. Therefore, even if there is  bulk superconductivity, the Meissner 
effect could be negligible because $\lambda_{\parallel} (0)/\bar{r}$ $>>$ 1. 
Only if the tubes are Josephson coupled in very low magnetic fields, 
the Meissner effect can be substantial.

It is known that the resistance of a superconducting wire measured 
through normal Ohmic contacts (two probe method) is not negligible because the number of the 
conductance channels in the wire is much smaller than in the contacts, 
as shown recently by  Kociak {\em et al.} \cite{Kociak}. 
Kociak {\em et al.} \cite{Kociak} observed 
superconductivity at 
0.55 K in single-wall carbon nanotube bundles; the resistance drops by two 
orders of magnitude below 0.55 K. Below $T_{c}$, a finite 
resistance of 74~$\Omega$ is observed for a bundle consisting of 350 
tubes in parallel  (the contact resistance is very small). This implies that 
the resistance of each tube below $T_{c}$ is equal 
to 74$\times$ 350 = 25.9 k$\Omega$ $\simeq$ 2$R_{Q}$ (where $R_{Q} = 
h/2e^{2}$ = 12.9 k$\Omega$). There have been no reports that this quantum 
resistance is observed in the normal state of any metallic single-wall nanotube with a 
length of 1~$\mu$m although a resistance of $R_{Q}/2$ is predicted for  
a normal metallic single-wall nanotube \cite{Chico} assuming ballistic 
transport. To 
ensure ballistic transport within a length of 1 
$\mu$m, the mean free path of a 
nanotube must be larger than 1 
$\mu$m, which is very unlikely at room temperature.  The inelastic scattering by a very 
low optical phonon mode (about 2 meV \cite{Saito}) should be rather strong at room 
temperature \cite{Pok}. Moreover, it was also shown that the linear $T$ dependence 
of the resistivity in metallic single-wall nanotubes arises from the 
inelastic scattering by a deformation mode (``twiston'') \cite{Kane}. These 
inelastic scatterings would strongly limit the mean free path at room 
temperature (the mean free path at room temperature can be estimated 
to be on the order of 100 \AA~from the measured resistivity), and make it 
impossible for a 1~$\mu$m long tube to have ballistic transport  at room 
temperature. 

In contrast, the quantum resistance of 
2$R_{Q}$ or $R_{Q}$ was observed for a 4~$\mu$m long multiwall nanotube 
even at room temperature \cite{Frank}. This was taken as evidence for 
ballistic transport in this multiwall nanotube \cite{Frank}. However, there have 
been no reports 
that  the quantum resistance is observed in the normal state of any 
metallic single-wall nanotubes.  The fact that the quantum resistance is observed 
only in the superconducting state of  a single-wall 
nanotube \cite{Kociak} implies that the observation of the 
quantum resistance should be associated with superconductivity in the tube. 
By analogy, the observation of the 
quantum resistance at room temperature in the multiwall nanotube  
\cite{Frank} implies that the multiwall nanotube is actually a room temperature 
superconductor. If  the superconductivity only occurs in the outer two layers of the 
tubes, one can naturally explain the observed quantum resistance of 
2$R_{Q}$  or $R_{Q}$ \cite{Frank}. The resistance of 2$R_{Q}$ corresponds to the 
case where only one of the superconducting layers is connected to the 
metal contacts, as in the case of the superconductivity in the single-wall nanotube \cite{Kociak}, while the resistance of $R_{Q}$ corresponds to the 
case where the two superconducting layers  are connected to the 
metal contacts. Furthermore, the tube does not dissipate heat with 
current density $> 10^{7}$ A/cm$^{2}$ \cite{Frank}. This is possible 
only if the tube is a room temperature superconductor or there is no inelastic 
scattering in the normal state of the tube. Since the inelastic 
scattering is inevitable at room temperature  \cite{Pok,Kane}, the 
nondissipative feature of the tube should be related to the superconductivity. 
Moreover, superconductivity of about 20 K has been observed in the 
metallic single-wall nanotubes \cite{Tang}. Presumely, the pairing 
mechanism in the single-wall nanotubes should be similar to that in
doped C$_{60}$, which is phonon-mediated \cite{Gunnarsson}. This 
implies that electrons should be 
coupled to lattice rather strongly in these nanotube materials so that the inelastic 
scattering is not very weak.

The much lower superconductivity in single-wall 
nanotubes than in multiwall nanotubes may be associated with the interlayer 
coupling strength. For multiwall nanotubes, the interlayer coupling 
should be much stronger than for the single-wall 
nanotubes, leading to a much higher $T_{c}$.  Theoretically, within the 
phonon-mediated mechanism, it is possible to have high temperature superconductivity in these carbon-based materials due to a very high 
phonon frequency ($\omega$ =2400 K) \cite{Saito}. A simple estimate suggests
that an electron-phonon coupling 
constant of  about 2 can lead to a $T_{c}$ of about 460 K. It was also 
shown that the very low-energy whispering mode can produce a very large 
pairing potential \cite{Pok}. Alexandrov and Mott \cite{Alex} estimated that the 
highest $T_{c}$ within a strong electron-phonon coupling model is about 
$\omega/3$, which is about 800 K.

 Although the present data along with the nondissipative feature and 
 the quantum resistance observed for the multiwall nanotubes 
 at room temperature \cite{Frank} can 
 be well explained by superconductivity above room temperature, we 
 cannot completely rule out the other possible explanations. For 
 example, ring currents around the tube axis could cause a diamagnetism 
 \cite{Ramirez}. Within this model, it seems difficult to explain why 
 the diamagnetic signal in 20 mG for the coupled tubes is about one order of 
 magnitude larger than for the physically separated tubes, and why 
 the diamagnetism in the low fields is so strongly dependent on the 
 field. Since the energy scale of the magnetic field is so small compared 
 with any other energy scales, it is hard to imagine 
 that the extreme sensitivity of 
 the diamagnetism to the field could be also understood by the other 
 models.  More rigorous theoretical and experimental efforts are 
 required to finally pin down whether the observed phenomena can be 
 only explained by superconductivity. Instead of  
 giving a define claim for the 
 superconductivity above room temperature in the multwall nanotubes, 
 we would like the readers to make their own judgment based on the 
 present experimental data.
~\\
 ~\\
 ~\\
 {\bf Acknowledgment:} We are grateful to R. L. 
 Meng for the source of the samples.  

~\\
~\\
* Correspondence should be addressed to gmzhao@uh.edu.

\end{document}